\documentclass[10pt,conference]{IEEEtran}
\usepackage{graphicx}
\usepackage{amsmath, amssymb}
\usepackage{bbm}
\usepackage{cite}
\usepackage{mathtools}
\usepackage[font=small]{caption}
\usepackage{subcaption}
\usepackage{bbold}
\usepackage{dsfont}
\usepackage{amsfonts}
\usepackage{tikz}
\usepackage{enumerate}
\usepackage{booktabs}
\usepackage{xcolor}

\usetikzlibrary{arrows,automata,positioning}

\usepackage{xcolor}

\title{MULTI-SCOUT: Multistatic Integrated Sensing and Communications in 5G and Beyond for Moving Target Detection, Positioning, and Tracking}

\author{
  Yalin E. Sagduyu\IEEEauthorrefmark{1}, \hspace{-0.025cm}Kemal Davaslioglu\IEEEauthorrefmark{1}, \hspace{-0.025cm}Tugba Erpek\IEEEauthorrefmark{1}, \hspace{-0.025cm}Sastry Kompella\IEEEauthorrefmark{1}, \hspace{-0.025cm}Gustave Anderson\IEEEauthorrefmark{2}, \hspace{-0.025cm}Jonathan Ashdown\IEEEauthorrefmark{3}\\
  \IEEEauthorrefmark{1}Nexcepta Inc., Gaithersburg, MD, USA\\
  \IEEEauthorrefmark{2}Lockheed Martin, Cherry Hill, NJ, USA\\
  \IEEEauthorrefmark{3}Air Force Research Laboratory, Rome, NY, USA
}

\date{}
\IEEEoverridecommandlockouts
\begin{document}

\maketitle
\thispagestyle{empty}

\begin{abstract}
This paper presents a complete signal-processing chain for multistatic integrated sensing and communications (ISAC) using 5G Positioning Reference Signal (PRS). We consider a distributed architecture in which one gNB transmits a periodic OFDM–PRS waveform while multiple spatially separated receivers exploit the same signal for target detection, parameter estimation and tracking. A coherent cross-ambiguity function (CAF) is evaluated to form a range–Doppler map from which the bistatic delay and radial velocity are extracted for every target. For a single target, bistatic delays are fused through nonlinear least-squares trilateration, yielding a geometric position estimate, and a regularized linear inversion of the radial-speed equations yields a two-dimensional velocity vector, where speed and heading are obtained. The approach is applied to 2D and 3D settings, extended to account for receiver clock synchronization bias, and generalized to multiple targets by resolving target association. The sequence of position–velocity estimates is then fed to standard and extended Kalman filters to obtain smoothed tracks. Our results show high-fidelity moving-target detection, positioning, and tracking using 5G PRS signals for multistatic ISAC.
\end{abstract}


\section{Introduction}
Integrated sensing and communications (ISAC) has been widely recognized as a key functionality for 5G and beyond systems \cite{liu2022integrated, liu2022survey, cui2021integrating, 9724260}. ISAC merges radar-style environment perception with traditional data delivery on a common radio-access infrastructure. Exploiting the existing cellular downlink for radar-like sensing avoids additional spectrum usage and hardware duplication, while providing native context awareness to support various vehicular, industrial, and security applications. In the tactical domain, ISAC supports Intelligence, Surveillance, and Reconnaissance (ISR) by enabling distributed units to share real-time battlefield intelligence with low latency and stealth through sensing–communication fusion.

ISAC systems vary by how the transmitter (illuminator) and receivers (sensors) are deployed: monostatic ISAC uses a co-located transmitter/receiver, bistatic ISAC separates them at two sites, and multistatic ISAC employs multiple geographically distributed receivers. While monostatic and bistatic configurations have been studied extensively \cite{iscout24,wei20225g,samczynski20215g,Wypich25,nataraja2024integrated, sagduyu2023joint, 
sagduyu2024will}, multistatic ISAC remains relatively unexplored, with only a handful of recent works such as \cite{Khosroshahi25}. Multistatic designs offer robustness to scattering geometry and blockage through diversity in range and Doppler observations, but also introduce challenges in extracting weak echoes, synchronizing time, associating targets across channels, and real-time tracking.

We address these challenges in the unified framework of MULTI-SCOUT for  multistatic ISAC with distributed receivers for moving target detection, positioning, and tracking. Using 5G positioning reference signal (PRS), the novel technical contributions of MULTI-SCOUT are: (i) blind multistatic range and Doppler estimation of moving targets, (ii) 2D/3D positioning via fused bistatic measurements, (iii) time synchronization, (iv) inter-receiver target association, and (v) tracking with Kalman filters (KFs).

We begin by building a multistatic estimator that pulls delay and Doppler peaks from each of three receivers' cross-ambiguity function (CAF) and fuses them into highly accurate bistatic range, velocity, and 2D position estimates. We then incorporate a blind clock-bias estimator via an auxiliary receiver to remove synchronization offsets. Extending this approach to non-coplanar sensor geometries with four receivers demonstrates that full 3D positioning can also achieve high accuracy. For the multi-target scenarios, we introduce a joint association scheme that pairs detections of targets across receivers and preserves high fidelity estimates even when echoes overlap. Finally, we show that applying KF to the fused estimates can produce smooth, low-error tracks. The Extended Kalman Filter (EKF) adapts to both linear and curved motions, while the standard (linear) KF is effective only for linear trajectories.

The remainder of the paper is organized as follows. Sec.~\ref{sec:single} describes single-target estimation. Secs.~\ref{sec:synch}, \ref{sec:3D}, and \ref{sec:multi} present time synchronization, 3D, and multi-target extensions. Sec.~\ref{sec:tracking} describes the tracking approach.  Sec.~\ref{sec:conclusion} concludes the paper.

\section{Single-Target Multistatic Estimation \label{sec:single}}
\begin{figure}[t!]
	\centering
	\includegraphics[width=0.76\columnwidth]{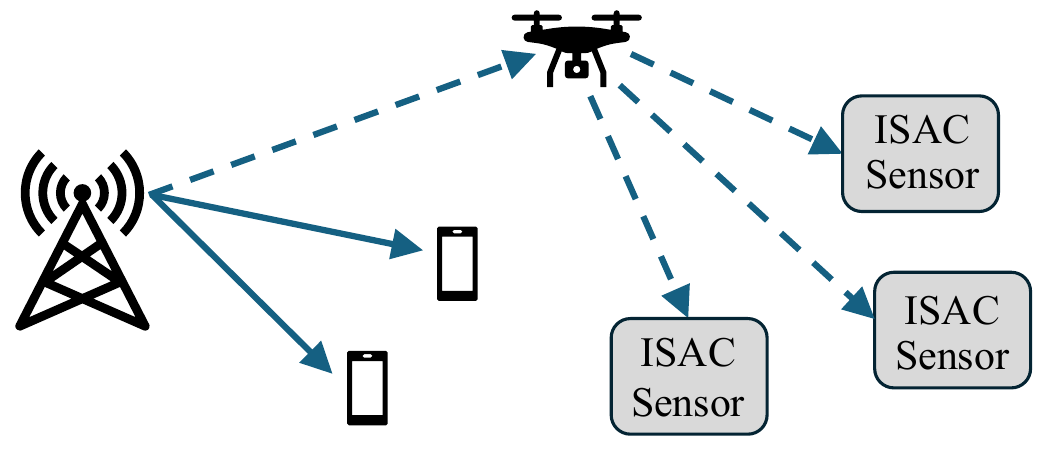}
	\caption{Multistatic ISAC setting.}
	\label{fig:system}
    \vspace{-0.1cm}
\end{figure}

We consider the trilateration setup in Fig.~\ref{fig:system}. Let $M$ denote the total number of receivers, indexed by $m=1,\ldots,M$. Each receiver $m$ measures its bistatic range to a target using the signal received from transmitter (5G gNB). Each measurement in 2D multistatic localization constrains the target to an ellipse with foci at the transmitter and a receiver. Two such ellipses may intersect ambiguously at up to four points, whereas a third independent ellipse produces a common intersection that resolves the ambiguity and pins down the unique location. Thus, 2D position estimation needs three non-collinear receivers.
\subsection{Signal model}
We consider 5G NR waveform of numerology~0 (subcarrier spacing $\Delta f = 15$~kHz) using orthogonal frequency-division multiplexing (OFDM).  The transmit timeline is formed by concatenating PRS symbols mapped on the FFT grid of length $N=1024$ with normal cyclic prefix (CP). We set the carrier frequency $f_c=2.5\,\text{GHz}$ and sampling rate $f_s=N\Delta f=15.36\,\text{MHz}$. PRS covers $K_{\mathrm{RB}}=76$ resource blocks centered within a Point-A--referenced carrier grid. The PRS is generated from a length-31 Gold-family sequence with initialization parameter $c_{\text{init}}$ and QPSK mapping in accordance with 3GPP TS~38.211 \cite{3gpp.38.211}, and mapped onto selected subcarriers within the Bandwidth Part (BWP) using a comb pattern of size $K_{\mathrm{comb}}=2$ and offset $k_{\mathrm{offset}}=0$, excluding the DC subcarrier. In each 14-symbol NR slot, we use $L_{\mathrm{PRS}}=2$ PRS symbols per slot. This structure repeats over slots to yield $K=128$ PRS symbols. Each frequency-domain PRS symbol is converted to time domain by a 1024-point IFFT and appended with the normal CP lengths specified in \cite{3gpp.38.211}.

A single point target is at Cartesian position $\mathbf{p}$ with velocity $\mathbf{v}$.  For the $m$-th receiver located at $\mathbf{r}_m$ the bistatic distance is
\begin{equation}
B_m(\mathbf{p})=\|\mathbf{p}-\mathbf{t}\|+\|\mathbf{p}-\mathbf{r}_m\|
\label{eq:B}
\end{equation}
for transmitter position  $\mathbf{t}$. The corresponding radial velocity is
\begin{equation}
v_m(\mathbf{p},\mathbf{v})=\mathbf{v}^\top
\left(\frac{\mathbf{p}-\mathbf{t}}{\|\mathbf{p}-\mathbf{t}\|}
      +\frac{\mathbf{p}-\mathbf{r}_m}{\|\mathbf{p}-\mathbf{r}_m\|}\right).
\label{eq:v}
\end{equation}

At each receiver, if present, direct-path echoes are canceled by matched-filtering the received samples with the PRS to estimate delay and gain, then subtracting the scaled, time-shifted replica.  The discretized baseband echo is given by
\begin{equation}
y_m[n]=\alpha_m\,s[n-\tau_m]\,
        \mathrm{e}^{\,\mathrm{j}2\pi f_{d,m} n T_{\mathrm{samp}}}+w_m[n], 
\label{eq:y}
\end{equation}
at receiver \(m\), where \(s[\cdot]\) is the complex baseband transmit signal, 
\(\tau_m\) is the sample-domain delay from the bistatic range, 
\(f_{d,m}=v_m f_c/c\), \(T_{\mathrm{samp}}=1/f_s\), 
\(w_m[n]\) is additive white Gaussian noise. The gain \(\alpha_m\)  captures propagation attenuation, phase, antenna gains, carrier wavelength, and target radar cross-section under the narrowband approximation. The coarse delay and Doppler are then refined by parabolic interpolation.

\subsection{CAF evaluation and peak interpolation}
For each receiver $m$, let $k=0,\dots,K-1$ index the $K$ PRS symbols, $t_k$ the mid-time (in seconds) of PRS symbol $k$ in the NR slot timeline, 
and $s_k[n]$ the CP-free PRS reference signal of length $N$ samples ($n=0,\dots,N-1$), derived from the corresponding PRS symbol within $s[n]$. For a delay grid $d\in\{0,\dots,D-1\}$  of $D$ bins and a Doppler grid of $N_f=401$ bins, we form the per-PRS matched-filter outputs
\begin{equation}
C_m(k,d)\;=\;\sum_{n=0}^{N-1} y_m[n_k^{\mathrm{cp}}+d+n]\;s_k^{*}[n],
\end{equation}
where $n_k^{\mathrm{cp}}$ is the sample index of the post-CP part of PRS symbol $k$.  Applying a Hann slow-time window $W[k]$ (of length $K$), the CAF is assembled via a non-uniform Discrete Fourier Transform (DFT) over the PRS mid-times on the discrete Doppler grid $f\in[-400,400]$ Hz:
\begin{equation}
\mathrm{CAF}_m(d,f)\;=\;\sum_{k=0}^{K-1} W[k]\;C_m(k,d)\; \mathrm{e}^{-\mathrm{j}2\pi f\, t_k}.
\label{eq:CAF}
\end{equation}
The magnitude $|\mathrm{CAF}_m(d,f)|$ gives the range–Doppler map. The global maximum $(d_0,f_0)$ yields coarse estimates $\hat B_m=d_0 c/f_s$ and $\hat v_m=f_0 c/f_c$. Parabolic (three-point) interpolation around the strongest bin refines the peak location, yielding delay and Doppler estimates with sub-sample precision.

\subsection{Position and velocity reconstruction}
Stacking the three bistatic-length equations in \eqref{eq:B} yields a nonlinear least-squares problem for synchronized receivers, 
\begin{equation}
\hat{\mathbf p} = \arg\min_{\mathbf p} \sum_{m=1}^{M} \rho_m(\mathbf p)^2,
\label{eq:ls}
\end{equation} 
where the residual error for receiver $m=1,\dots,M$ is
\begin{equation}
\rho_m(\mathbf p)
=\|\mathbf p-\mathbf t\| + \|\mathbf p-\mathbf r_m\| - \hat B_m.
\end{equation}

For $M=3$, the problem is solved by a Trust-Region Reflective (TRF) nonlinear least-squares solver with random initializations to avoid local minima. Each iteration computes a damped normal–equations step 
\[
  \bigl(\mathbf{J}^\top \mathbf{J} + \eta\,\mathbf{I}\bigr)\,\delta\mathbf{p}
  = -\mathbf{J}^\top \boldsymbol{\rho},
\]
where \(\boldsymbol{\rho}=[\rho_1(\mathbf p),\ldots,\rho_M(\mathbf p)]^\top\), \(\mathbf{J}\) is its Jacobian, and \(\eta\) interpolates between Gauss–Newton (\(\eta\to0\)) and gradient descent (\(\eta\) large). 

Because the cost surface may contain local minima, we draw initial guesses uniformly from a box extending 200~m beyond the convex hull of \(\{\mathbf{t},\mathbf{r}_1,\mathbf{r}_2,\mathbf{r}_3\}\). For each seed, the solver is run to convergence, and the solution with the smallest residual is taken as \(\hat{\mathbf{p}}\). The final sum of squared residuals is the trilateration cost. Once $\hat{\mathbf{p}}$ is found, \eqref{eq:v} becomes a linear system in $\mathbf{v}$, solved by ridge-regularized inverse
\begin{equation}
\hat{\mathbf{v}}=
\left( \mathbf{A}^\top\mathbf{A}+\varepsilon\mathbf{I}\right)^{-1}
\mathbf{A}^\top\hat{\mathbf{v}}_r,
\label{eq:vhat}
\end{equation}
where $\mathbf{A}\in\mathbb{R}^{3\times 2}$ contains the unit vectors of \eqref{eq:v} evaluated at $\hat{\mathbf{p}}$, and $\hat{\mathbf{v}}_r=[\hat v_1,\hat v_2,\hat v_3]^\top$ collects the CAF-estimated radial velocities $\hat v_m$ from each receiver. A ridge regression (Tikhonov regularization) term $\varepsilon\,\mathbf{I}$ is added ($\varepsilon = 10^{-3}$) to guard against ill‐conditioning of $\mathbf{A}^\top\mathbf{A}$, yielding the solution in \eqref{eq:vhat}. Speed and heading follow as $\|\hat{\mathbf{v}}\|$ and $\operatorname{atan2}(\hat v_y,\hat v_x)$, the two-argument arctangent giving angle (azimuth) in $(-\pi,\pi]$.

\subsection{Performance Evaluation}

We consider three receivers positioned on an equilateral triangle: $\mathbf{r}_1 = (0,0)$ m, $\mathbf{r}_2 = (500,0)$ m, and $\mathbf{r}_3 \approx (250,433)$ m. 
The transmitter is located at the centroid of this triangle, $\mathbf t \approx  (250, 144)$ m. For each target, the position is selected uniformly randomly from the area of $[0,500] \times [0,500]$, the speed is selected uniformly randomly between 20 and 30\,m/s and the angle is selected uniformly randomly in $[0,2\pi)$. Transmit power is 40~dBm (isotropic), transmit and receive antenna gains are 10~dBi, radar cross section is 4~m$^{2}$, and noise variance is   $10^{-3}$ (per complex baseband sample).

\begin{figure}[ht!]
	\centering
	\includegraphics[width=0.945\columnwidth]{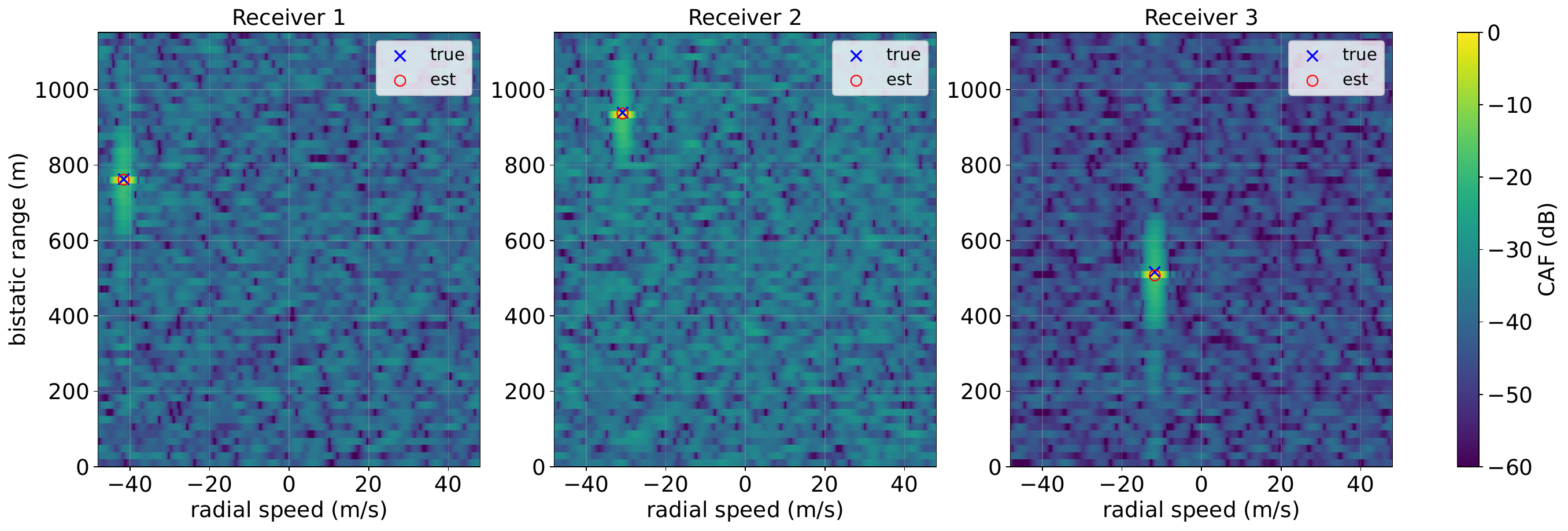}
	\caption{Range-Doppler maps for a single target.}
	\label{fig:singletarget_map}
    \vspace{-0.4cm}
\end{figure}

\begin{figure}[ht!]
	\centering
	\includegraphics[width=0.6\columnwidth]{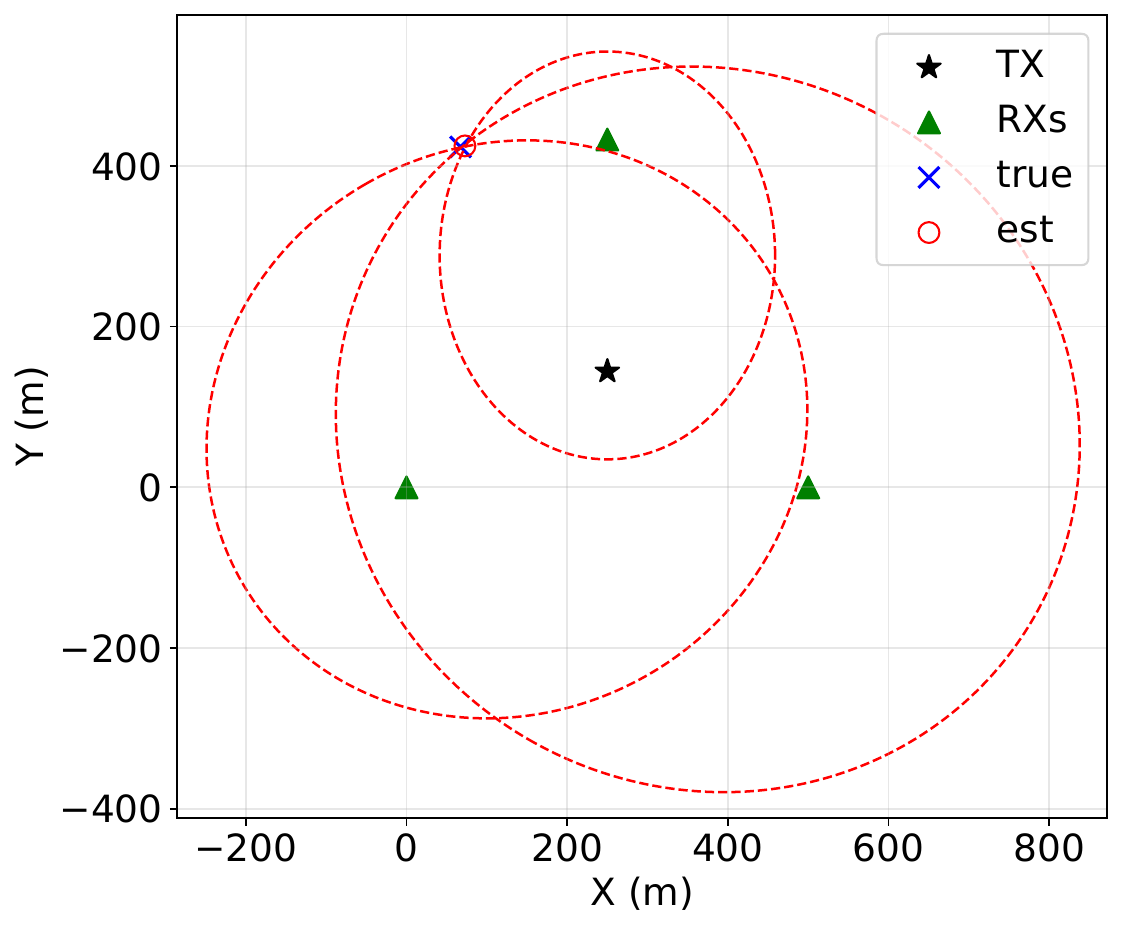}
	\caption{Bistatic ellipses and trilateration for a single target.}
	\label{fig:singletarget_position}
\end{figure}

\begin{figure}[ht!]
	\centering
	\includegraphics[width=0.45\columnwidth]{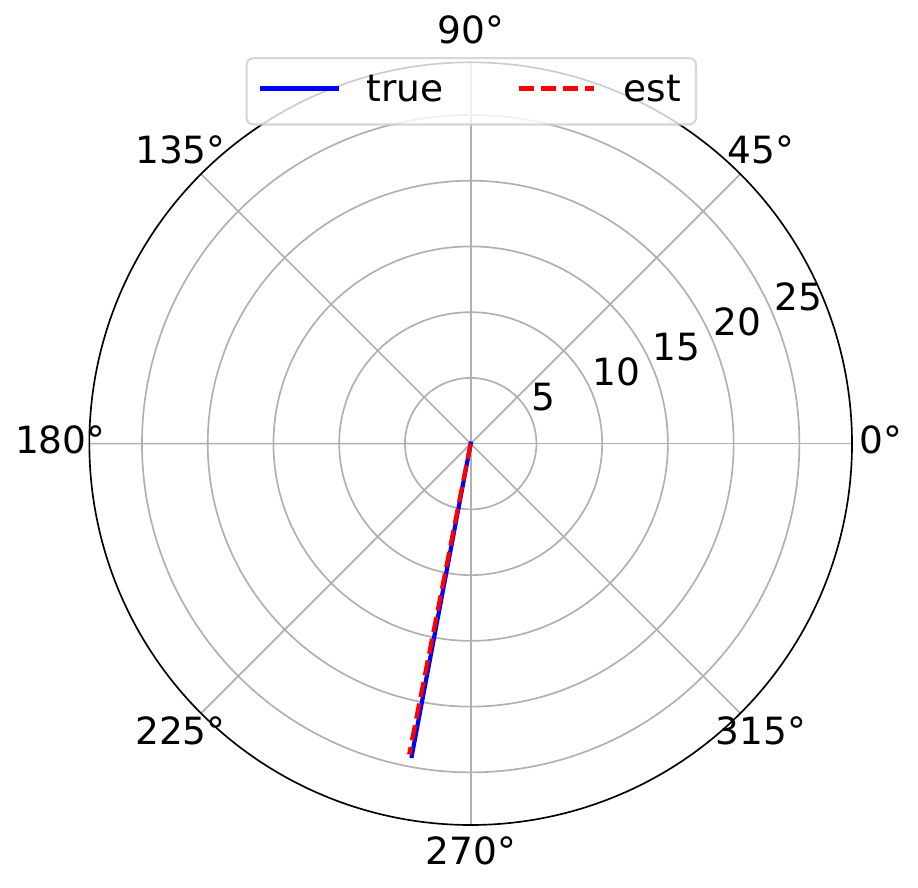}
	\caption{Velocity vector (polar) for a single target.}
	\label{fig:singletarget_velocity}
\end{figure}

Fig.~\ref{fig:singletarget_map} shows the Range-Doppler maps at three different receivers. 
For each receiver $m$, Table~\ref{tab:per_rx_bistatic} shows highly accurate per‐receiver bistatic estimates of range $\hat B_m$ and velocity $\hat v_m$ relative to their ground‐truth values $B_m$ and $v_m$ across all receivers.
The resulting bistatic ellipses and trilateration are visualized in Fig.~\ref{fig:singletarget_position}. The errors in position, speed, and angle estimates are small, as shown in Table~\ref{tab:target_pos_vel}. Fig.~\ref{fig:singletarget_velocity} shows the estimated target velocity vector in polar coordinates by combining Doppler measurements across the receivers. We present the root-mean-square (RMS) bistatic-range error, absolute error in speed and angle, and overall position error in Euclidean distance in Table~\ref{tab:performance_metrics1}. For average statistics, the experiment is repeated 100 times using randomly generated target locations and the average performance results are presented in Table~\ref{tab:performance_metrics2}.

\begin{table}[htbp]
  \centering
  \caption{Per‐receiver bistatic estimates for a single target.}
  \label{tab:per_rx_bistatic}
  \vspace{-0.1cm}
  \begin{tabular}{ccccc}
    \toprule
    Receiver $m$& $B_m$ (m) & $\hat B_m$ (m) & $v_m$ (m/s) & $\hat v_m$ (m/s) \\
    \midrule
    1 & 762.89 & 761.69 & -41.58 & -41.60 \\
    2 & 939.58 & 937.22 & -30.83 & -30.85 \\
    3 & 516.94 & 507.66 & -11.74 & -11.74 \\
    \bottomrule
  \end{tabular}
\end{table}
\vspace{-0.4cm}

\begin{table}[htbp]
  \centering
  \caption{Position and velocity estimates for a single target.}
  \label{tab:target_pos_vel}
  \begin{tabular}{lll}
    \toprule
    & \multicolumn{1}{c}{True} & \multicolumn{1}{c}{Estimated} \\
    \midrule
    Position (x, y) (m)
      & $[67.18,\,423.72]$
      & $[72.71,\,424.99]$ \\
    Speed 
      & $24.17\text{\,m/s}$
      & $24.11\text{\,m/s}$  \\
    Angle & $-100.68^\circ$ & $-101.20^\circ$ \\
    \bottomrule
  \end{tabular}
\end{table}

\begin{table}[h!]
  \centering
  \caption{Performance for single target positioning.}
  \label{tab:performance_metrics1}
  \begin{tabular}{ll}
    \toprule
    Metric & Value \\
    \midrule
    Trilateration cost 
      & 12.02 \\
    RMS bistatic‐range error 
      & 5.57\,m  (0.75\% of mean bistatic range) \\
    Absolute error in speed
      & 0.07\,m/s (0.27\% of true speed) \\
    Absolute error in angle 
      & 0.51$^\circ$ (0.14\% of full circle) \\
     Position error  & 5.67\,m\\
    \bottomrule
  \end{tabular}
\end{table}

\begin{table}[h!]
  \centering
  \caption{Averaged results for single target positioning.}
  \label{tab:performance_metrics2}
  \begin{tabular}{ll}
    \toprule
    Metric & Value \\
    \midrule
    Trilateration cost 
      & 14.08 \\
    RMS bistatic‐range error 
      & 5.08\,m (0.98\% of mean bistatic range) \\
    Absolute error in speed  
      & 0.37\,m/s (1.53\% of true speed) \\
    Absolute error in angle 
      & 1.04$^\circ$ (0.29\% of full circle) \\
    Position error 
      & 5.37\,m \\
    \bottomrule
  \end{tabular}
\end{table}

\vspace{-0.2cm}
\section{Time Synchronization in Multistatic ISAC \label{sec:synch}}
In Sec.~\ref{sec:single}, when receiver clocks were perfectly synchronized, three receivers in 2D were sufficient to estimate the target position by solving (\ref{eq:ls}).
If the receiver clocks instead share an unknown common offset \(\delta_t\), each bistatic-range measurement acquires a bias \(c\,\delta_t\). With $M \geq 3$ receivers, we can estimate $(\hat{\mathbf p},\widehat{\delta_t})
= \arg\min_{\mathbf p,\delta_t}
\sum_{m=1}^{M} \rho_m(\mathbf p,\delta_t)^2$, where the residual error for receiver $m$ is given by
\begin{align}
\rho_m(\mathbf p,\delta_t)
=\|\mathbf p-\mathbf t\| + \|\mathbf p-\mathbf{r}_m\|
-\bigl(\hat B_m + c\,\delta_t\bigr)
\end{align}
for $m=1,\dots,M$.  For $M =3$, this system with three equations and three unknowns can yield a unique solution provided the three receivers are not collinear. By introducing a fourth receiver ($M=4$), we obtain an over-determined system to estimate $\hat{\mathbf p}$. The extra measurement improves the conditioning and robustness of the joint estimation of the clock bias and position. The first three receivers are positioned as before and the fourth receiver's position is (0,500) m. We set $\delta_t = 20$~ns, corresponding to a common 
range bias of about $6$~m. Table~\ref{tab:per_rx_bistatic_time} shows the bistatic estimates, Table~\ref{tab:target_pos_vel_time} shows the target position and velocity estimates, and Table~\ref{tab:performance_metrics_time} shows the target positioning performance. Overall, estimates remain reliable, and adding a fourth receiver improves robustness to the clock bias. For the rest of the paper, we continue with the assumption of synchronized receivers.

\begin{table}[htbp]
  \centering
  \caption{Bistatic estimates with time synchronization.}
  \label{tab:per_rx_bistatic_time}
  \begin{tabular}{ccccc}
    \toprule
    Receiver $m$& $B_m$ (m) & $\hat B_m$ (m) & $v_m$ (m/s) & $\hat v_m$ (m/s) \\
    \midrule
    1 & 762.89 & 761.66 & -41.58 & -41.57 \\
    2 & 939.58 & 937.35 & -30.83 & -30.89 \\
    3 & 516.94 & 527.18 & -11.74 & -11.75 \\
    4 & 435.53 & 449.08 &  -2.56 &  -2.55 \\
    \bottomrule
  \end{tabular}
\end{table}

\begin{table}[htbp]
  \centering
  \caption{Time synchronization effects on target estimates.}
  \label{tab:target_pos_vel_time}
  \begin{tabular}{lll}
    \toprule
    & \multicolumn{1}{c}{True} & \multicolumn{1}{c}{Estimated} \\
    \midrule
    Position (x, y) (m)
      & $[67.18,\,423.72]$
      & $[66.58,\,413.84]$ \\
    Speed 
      & $24.17\text{\,m/s}$
      & $23.92\text{\,m/s}$  \\
    Angle & $-100.68^\circ$ & $-98.70^\circ$ \\
    \bottomrule
  \end{tabular}
\end{table}

\begin{table}[h!]
  \centering
  \caption{Positioning with time synchronization.}
  \label{tab:performance_metrics_time}
  \begin{tabular}{ll}
    \toprule
    Metric & Value \\
    \midrule
    Trilateration cost
      & 14.47 \\
    RMS bistatic‐range error 
      & 6.98\,m  (1.04\% of mean bistatic range) \\
    Absolute error in speed
      & 0.25\,m/s (1.03\% of true speed) \\
    Absolute error in angle 
      & 1.98$^\circ$ (0.55\% of full circle) \\
    Position error 
      & 9.89\,m \\
    \bottomrule
  \end{tabular}
\end{table}

\section{3D Extension of Multistatic ISAC \label{sec:3D}}

In 3D, we recover $\mathbf{p}=[x,y,z]^\top$, so an additional bistatic length equation is required. Stacking four measurements yields
\begin{equation}
\hat{\mathbf{p}}
=\operatorname*{arg\,min}_{\mathbf{p}}
\sum_{m=1}^{4}\Bigl(
  \|\mathbf{p}-\mathbf{t}\|+\|\mathbf{p}-\mathbf{r}_m\|-\hat B_m
\Bigr)^2,
\label{eq:ls3d}
\end{equation}
which is solved via nonlinear least squares. The Jacobian $\mathbf{J}\in\mathbb{R}^{4\times3}$ is full rank only if the four receivers $\{\mathbf{r}_m\}$ are not coplanar. We stack the Doppler measurements to form a $4\times3$ matrix $\mathbf{A}$, whose $m$th row is the sum of the unit vectors from $\hat{\mathbf p}$ to the transmitter and to receiver $m$. The velocity is recovered by the same ridge-regularized least squares solution
\begin{equation}
\hat{\mathbf v}
=\bigl(\mathbf{A}^\top\mathbf{A}+\varepsilon\,\mathbf{I}\bigr)^{-1}
\mathbf{A}^\top\hat{\mathbf v}_r,
\end{equation}
exactly as in 2D but with $\mathbf{A}\in\mathbb R^{4\times3}$. As in 2D, speed and direction (azimuth) follow from $\|\hat{\mathbf{v}}\|$ and $\operatorname{atan2}(\hat v_y,\hat v_x)$.

Four receivers are located at positions $(0,0,0)$, $(500,0,0)$, $(0,500,0)$, and $(0,0,500)$~m forming a tetrahedron, while the transmitter is located at their centroid at $(125,125,125)$~m.
Table~\ref{tab:3Dper_rx_bistatic} presents the bistatic estimates and Table~\ref{tab:3Dtarget_pos_vel} presents the target position and velocity estimates, when the four sensors are combined. Table~\ref{tab:3Dperformance_metrics} shows the target positioning performance. Overall, positioning error remains small in 3D. For the rest of the paper, we continue with the 2D setting.

\begin{table}[htbp]
  \centering
  \caption{Per‐receiver bistatic estimates in 3D.}
  \label{tab:3Dper_rx_bistatic}
  \begin{tabular}{ccccc}
    \toprule
    Receiver $m$& $B_m$ (m) & $\hat B_m$ (m) & $v_m$ (m/s) & $\hat v_m$ (m/s) \\
    \midrule
    1 & 972.56  & 976.35  & -35.19 & -35.20 \\
    2 & 1114.24 & 1112.98 & -21.92 & -21.91 \\
    3 & 793.39  & 800.62  & -34.49 & -34.49 \\
    4 & 843.18  & 839.80  & -19.98 & -19.97 \\
    \bottomrule
  \end{tabular}
\end{table}

\begin{table}[htbp]
  \centering
  \caption{Target position and velocity estimates in 3D.}
  \label{tab:3Dtarget_pos_vel}
  \begin{tabular}{lll}
    \toprule
    & \multicolumn{1}{c}{True} & \multicolumn{1}{c}{Estimated} \\
    \midrule
    Position (m)
      & $[67.18,\,423.72,\,381.89]$
      & $[72.47,\,420.66,\,387.61]$ \\
    Speed 
      & $24.17\text{\,m/s}$
      & $23.98\text{\,m/s}$ \\
    Angle 
      & $-150.77^\circ$
      & $-152.00^\circ$ \\
    \bottomrule
  \end{tabular}
\end{table}

\begin{table}[h!]
  \centering
  \caption{Performance for target positioning in 3D.}
  \label{tab:3Dperformance_metrics}
  \begin{tabular}{ll}
    \toprule
    Metric & Value \\
    \midrule
    Trilateration cost 
      & 1.28 \\
    RMS bistatic‐range error 
      & 4.46\,m (0.48\% of mean bistatic range) \\
    Speed error
      & 0.19\,m/s (0.77\% of true speed) \\
    Angle error
      & 1.23$^\circ$ (0.34\% of full circle) \\
    Position error 
      & 8.37\,m \\
    \bottomrule
  \end{tabular}
\end{table}

\section{Multi-Target Multistatic Estimation \label{sec:multi}}
We consider $N_T>1$ simultaneous targets. Each CAF map $\mathrm{CAF}_m(d,f)$
shows up to $N_T$ peaks; taking only the strongest can lock onto sidelobes or clutter.
Instead, per receiver we apply 2D non-maximum suppression (NMS) on
$\lvert \mathrm{CAF}_m(d,f)\rvert$ with guard windows in delay and Doppler to select
exactly $N_T$ distinct delay rows. For each selected peak, parabolic (three-point)
interpolation in Doppler and delay refines sub-bin Doppler frequency and bistatic range.
Each receiver returns an ordered (by magnitude) list of $N_T$ delay rows with indices
$(d_{m,0},\dots,d_{m,N_T-1})$. A global assignment is an $M$-tuple of permutations
$(\pi_1,\dots,\pi_M)$ (here $M=3$) mapping target $k$ to the delay tuple
$\bigl(d_{1,\pi_1(k)},\,d_{2,\pi_2(k)},\,d_{3,\pi_3(k)}\bigr)$; the best assignment
minimizes the sum of squared trilateration residuals. For any assignment hypothesis, we solve \eqref{eq:ls} per target with the associated
(refined) delays; the objective value of \eqref{eq:ls} is the target cost $c_k$
(multi-start, keep the smallest residual). The total cost is $C=\sum_{k=0}^{N_T-1} c_k$,
and the hypothesis with minimal $C$ yields the index mapping and positions. The search
space has $(N_T!)^{M}$ hypotheses (e.g., $8$ if $M=3$ and $N_T=2$); for larger $N_T$,
branch-and-bound can prune the search space. After assignment, the matched Dopplers are converted to
radial speeds (forming $\hat{\mathbf v}_r$ by stacking one per receiver for each target)
and \eqref{eq:vhat} is applied per target to recover full velocity vectors.

\begin{figure}[ht!]
\vspace{-0.1cm}
	\centering
	\includegraphics[width=\columnwidth]{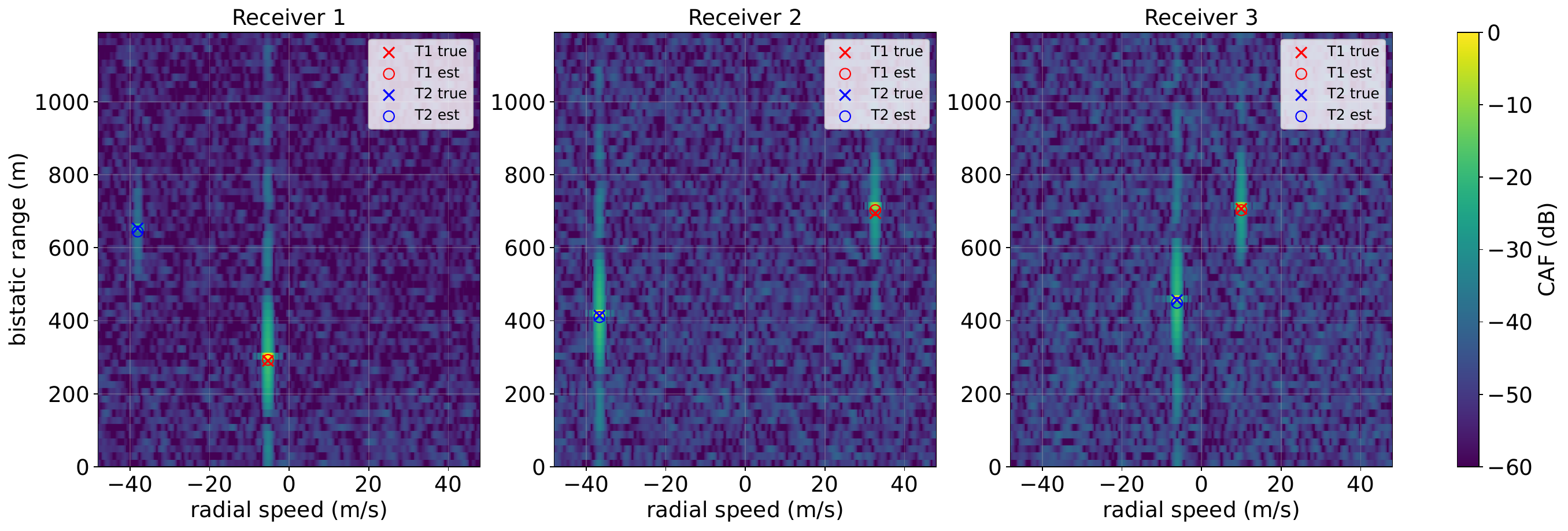}
	\caption{Range-Doppler maps for multiple targets (T1, T2).}
	\label{fig:multitarget_map}
\end{figure}

\begin{figure}[ht!]
\vspace{-0.3cm}
	\centering
	\includegraphics[width=0.6\columnwidth]{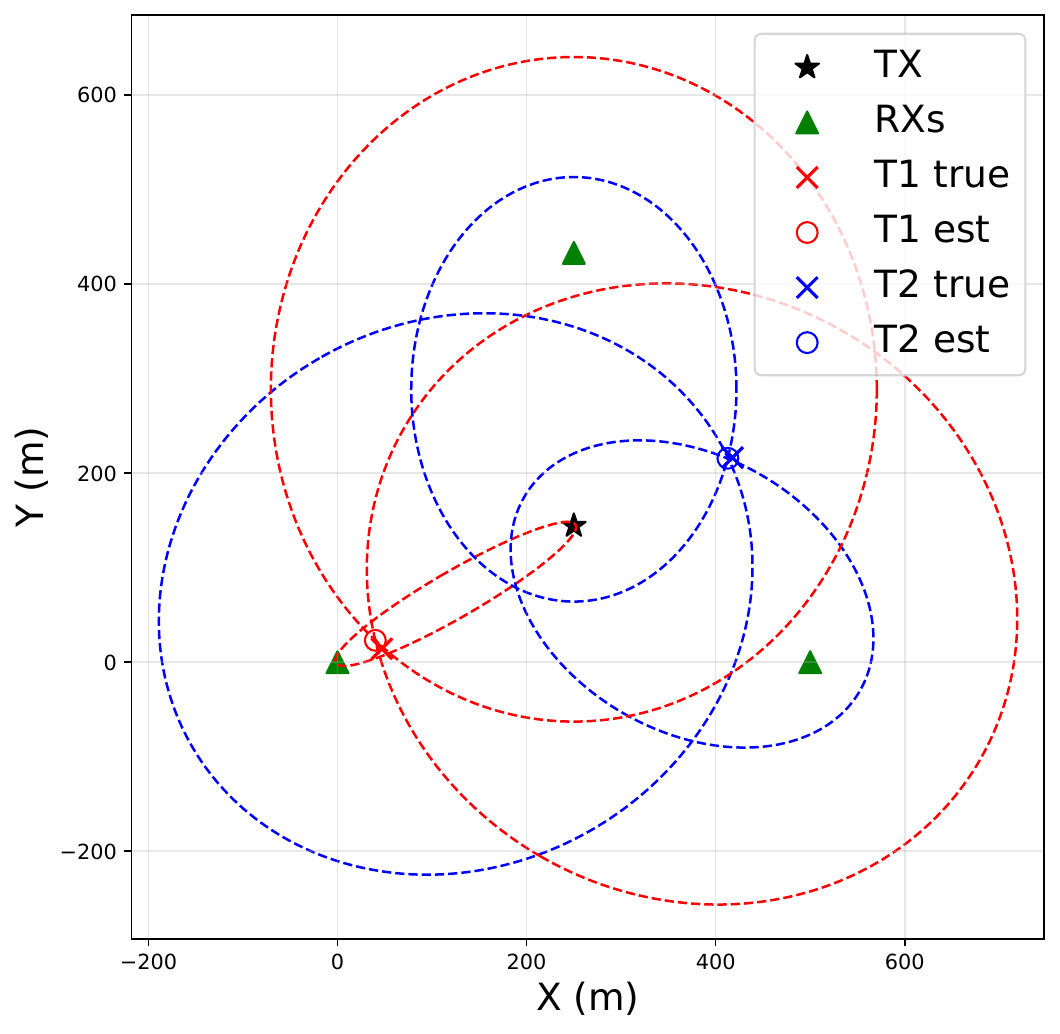}
	\caption{Bistatic ellipses and trilateration for multi-target case.}
	\label{fig:multitarget_position}
\end{figure}

\begin{figure}[ht!]
    \vspace{-0.5cm}
	\centering
	\includegraphics[width=0.45\columnwidth]{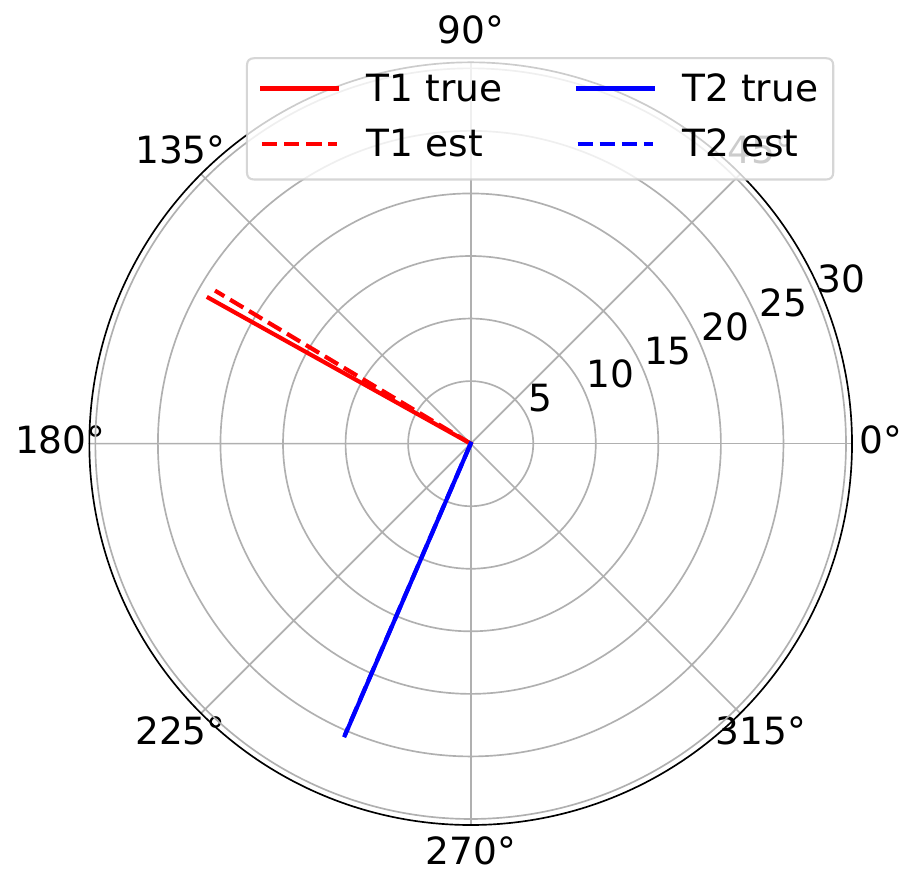}
	\caption{Velocity vector (polar) for multi-target case.}
	\label{fig:multitarget_velocity}
    \vspace{-0.1cm}
\end{figure}

Table~\ref{tab:blind_association_costs} evaluates eight possible triple-permutation hypotheses, revealing two symmetric permutations that result in the lowest trilateration cost and correctly identify the target association. For this association, Fig.~\ref{fig:multitarget_map} shows the Range-Doppler map at three receivers, Table~\ref{tab:per_rx_bistatic_estimates} presents the bistatic estimates per receiver and per target, Fig.~\ref{fig:multitarget_position} shows the bistatic ellipses and trilateration cost for the multi-target case,
Table~\ref{tab:matched_target_pos_vel} presents the multi-target position and velocity estimates, Fig.~\ref{fig:multitarget_velocity} shows the velocity vectors for targets, and  Table~\ref{tab:trilateration_metrics} presents the error performance. Considering randomly-selected target pair locations, Table~\ref{tab:multi_performance_metrics} shows the average performance. Results show that estimates are highly reliable with low errors when we consider multiple targets.

\begin{table}[htbp]
  \vspace{-0.2cm}
  \centering
  \caption{Target associations and their trilateration costs.}
  \label{tab:blind_association_costs}
  \begin{tabular}{ll}
    \toprule
    Candidate Target Pairs & Trilateration Cost \\
    \midrule
    $((0,1),(0,1),(0,1))$ & 44849.10\\
    $((0,1),(0,1),(1,0))$ & 22708.49 \\
    $((0,1),(1,0),(0,1))$ & 14960.25 \\
    $((0,1),(1,0),(1,0))$ & 17.95   \\
    $((1,0),(0,1),(0,1))$ & 17.95    \\
    $((1,0),(0,1),(1,0))$ & 14960.25  \\
    $((1,0),(1,0),(0,1))$ & 22708.49 \\
    $((1,0),(1,0),(1,0))$ & 44849.10 \\
    \bottomrule
  \end{tabular}
  \vspace{-0.1cm}
\end{table}

\begin{table}[htbp!]
  \centering
  \caption{Per‐receiver and per-target bistatic estimates.}
  \label{tab:per_rx_bistatic_estimates}
  {
  \setlength{\tabcolsep}{4pt} %
  \begin{tabular}{cccccc} 
    \toprule
    Receiver $m$ & Target & $B_m$ (m) & $\hat B_m$ (m) & $v_m$ (m/s) & $\hat v_m$ (m/s) \\
    \midrule
    1 & 1 & 290.23 & 292.82 & -5.34 & -5.34 \\
    1 & 2 & 653.27 & 644.34 & -38.08 & -38.07 \\
    2 & 1 & 694.50 & 703.01 & 32.64 & 32.64 \\
    2 & 2 & 414.13 & 410.01 & -36.66 & -36.66 \\
    3 & 1 & 706.66 & 702.98 & 10.00 & 10.00 \\
    3 & 2 & 456.75 & 449.08 & -6.16 & -6.16 \\
    \bottomrule
  \end{tabular}
  }
\end{table}

\begin{table}[htbp!]
  \centering
  \caption{Multi-target position and velocity estimates.}
  \label{tab:matched_target_pos_vel}
  \begin{tabular}{c l l l}
    \toprule
    Target & Parameter & True & Estimated \\
    \midrule
    1 & Position (m) & $[46.93,\,14.17]$ & $[40.02,\,23.13]$ \\
      & Speed (m/s)  & $23.97$           & $23.82$           \\
      & Angle (°)    & $150.91^\circ$    & $149.10^\circ$    \\
    \midrule
    2 & Position (m) & $[417.88,\,216.38]$ & $[412.97,\,215.69]$ \\
      & Speed (m/s)  & $25.39$             & $25.40$             \\
      & Angle (°)    & $-113.32^\circ$     & $-113.35^\circ$     \\
    \bottomrule
  \end{tabular}
\end{table}

\begin{table}[htbp!]
  \centering
  \caption{Performance for multi-target positioning.}
  \label{tab:trilateration_metrics}
  \begin{tabular}{ll}
    \toprule
    Metric & Value \\
    \midrule
    Trilateration cost 
      & 17.95 \\
    RMS bistatic‐range error
      & 6.43\,m (1.20 \% of mean bistatic range) \\
    Absolute error in speed
      & 0.10\,m/s (0.42\%) \\
    Absolute error in angle
      & 1.28$^\circ$ (0.36\% of full circle) \\
    Position error  & 8.14\,m\\
    \bottomrule
  \end{tabular}
\end{table}

\begin{table}[htbp!]
  \centering
  \caption{Averaged results for multi-target positioning.}
  \label{tab:multi_performance_metrics}
  \begin{tabular}{ll}
    \toprule
    Metric & Value \\
    \midrule
    Trilateration cost 
      & 19.54 \\
    RMS bistatic‐range error 
      & 5.43\,m (1.00\% of mean bistatic range) \\
    Absolute error in speed  
      & 0.49\,m/s (2.02\% of true speed) \\
    Absolute error in angle 
      & 3.08$^\circ$ (0.86\% of full circle) \\
     Position error  & 5.01\,m\\ 
    \bottomrule
  \end{tabular}
  \vspace{-0.2cm}
\end{table}

\section{Tracking with Kalman Filters} \label{sec:tracking}
After estimating the target positions and velocities, the next step is to track each target. The KF continuously blends incoming, noisy sensor readings with a prediction from a motion model to best estimate the target location and motion.

\subsection{Standard Kalman Filter}
The estimator of Sec.~\ref{sec:single} delivers noisy snapshots 
\(\mathbf{z}_k=[\hat x,\hat y,\hat v_x,\hat v_y]_k^\top\) at discrete instants \(kT\), where \(T\) is set now as the KF sampling period.  Assuming constant–velocity motion within each interval, the kinematic state evolves as
\begin{equation}
\mathbf{x}_{k+1}= \mathbf{F}\,\mathbf{x}_k+\mathbf{q}_k,
\quad
\mathbf{F} \;=\;
\begin{bmatrix}
\mathbf{I} & T\,\mathbf{I} \\[6pt]
\mathbf{0} & \mathbf{I}
\end{bmatrix},
\label{eq:F}
\end{equation}
where $\mathbf{I}$ is the identity matrix, $\mathbf{0}$ is the zero matrix, and  \(\mathbf{q}_k\sim\mathcal{N}(\mathbf{0},\mathbf{Q})\) is process noise.  
The measurement equation is identity,
\(\mathbf{z}_k=\mathbf{H}\mathbf{x}_k+\mathbf{n}_k\), 
\(\mathbf{H}=\mathbf{I}\), where \(\mathbf{n}_k\sim\mathcal{N}(\mathbf{0},\mathbf{R})\).  
The predict–update recursions are
\begin{equation}
\begin{aligned}
& \hat{\mathbf{x}}_{k|k-1} = \mathbf{F}\,\hat{\mathbf{x}}_{k-1|k-1}, \:
P_{k|k-1} = \mathbf{F}\,P_{k-1|k-1}\,\mathbf{F}^{\!\top} + \mathbf{Q}, \\
& K_k = P_{k|k-1}\,\mathbf{H}^{\!\top}\bigl(\mathbf{H}\,P_{k|k-1}\,\mathbf{H}^{\!\top} + \mathbf{R}\bigr)^{-1}, \\
& \hat{\mathbf{x}}_{k|k} = \hat{\mathbf{x}}_{k|k-1} + K_k\bigl(\mathbf{z}_k - \mathbf{H}\,\hat{\mathbf{x}}_{k|k-1}\bigr), \\
& P_{k|k} = (\mathbf{I} - K_k \mathbf{H})\,P_{k|k-1}.
\end{aligned}
\end{equation}

The standard KF relies on a predefined state-transition model 
$\mathbf{F}$ that must match the true target dynamics (e.g., constant velocity). When the target accelerates or turns, this model mismatch leads to degraded tracking performance.

\subsection{Extended Kalman Filter (EKF)}
Instead of fixed motion models, the state is augmented to
$\mathbf{x}_k = [x_k, y_k, v_k, \theta_k]^\top$ and the nonlinear process for prediction is
$f(\mathbf{x}) = [x + T\, v\cos\theta,\; y + T\, v\sin\theta,\; v,\; \theta]^\top$.
The measurement function maps the augmented state to the observed Cartesian components as
$h(\mathbf{x}) = [x,\; y,\; v\cos\theta,\; v\sin\theta]^\top$, and the actual measurement is
$\mathbf{z}_k = h(\mathbf{x}_k) + \mathbf{n}_k$ with $\mathbf{n}_k\sim\mathcal{N}(\mathbf{0},\mathbf{R})$.
We denote the EKF Jacobians by $\mathbf{J}_k^f$ for process model $f$ and $\mathbf{J}_k^h$ for measurement model $h$. Using these Jacobians (after linearization), the EKF recursions become
\begin{equation}
\begin{aligned}
&\hspace{-0.17cm}\hat{\mathbf{x}}_{k|k-1} \hspace{-0.04cm}= \hspace{-0.04cm}f(\hat{\mathbf{x}}_{k-1|k-1}), 
P_{k|k-1} \hspace{-0.04cm}= \hspace{-0.03cm}\mathbf{J}_k^f\,P_{k-1|k-1}\,(\mathbf{J}_k^f)^{\!\top} \hspace{-0.02cm} + \hspace{-0.02cm} \mathbf{Q}, \hspace{-0.08cm}\\
&\hspace{-0.17cm}K_k = P_{k|k-1}\,(\mathbf{J}_k^h)^{\!\top}\bigl(\mathbf{J}_k^h\,P_{k|k-1}\,(\mathbf{J}_k^h)^{\!\top} + \mathbf{R}\bigr)^{-1}, \\
&\hspace{-0.17cm}\hat{\mathbf{x}}_{k|k} = \hat{\mathbf{x}}_{k|k-1} + K_k\bigl(\mathbf{z}_k - h(\hat{\mathbf{x}}_{k|k-1})\bigr), \\
&\hspace{-0.17cm}P_{k|k} = (\mathbf{I} - K_k \mathbf{J}_k^h)\,P_{k|k-1}.
\end{aligned}
\end{equation}

By representing velocity in polar form \((v,\theta)\) but measuring Cartesian velocity components \((v_x,v_y)\), the EKF predicts \((x,y)\) along the current heading and updates \(v\) and \(\theta\) from the Cartesian velocity observations. Process noise on $(v,\theta)$ accounts for acceleration and turns, enabling the filter to operate without a fixed motion model and learn the trajectory from the \((x,y,v,\theta)\) sequence under smoothness constraints.

We consider tracking a target under linear and circular-type motions. The target proceeds at a steady nominal speed, modulated by small 10\% fluctuations to mimic throttle or wind variations. In circular motion, it performs a constant‐rate turn with a 25~sec period, tracing a circular path such that the true velocity rotates at constant angular rate, causing the KF to correct the occasional outliers produced by the trilateration stage. KF sampling period $T$ is 1~sec, tracking continues over 25~sec, and the KF noise covariances are $\mathbf{Q}=\operatorname{diag}(10^{-4},10^{-4},10^{-2},10^{-2})$, and $\mathbf{R}=\operatorname{diag}(10,10,1,1)$

\begin{figure}[ht!]
  \centering
  \includegraphics[
    width=0.7\columnwidth,
    height=0.65\columnwidth
  ]{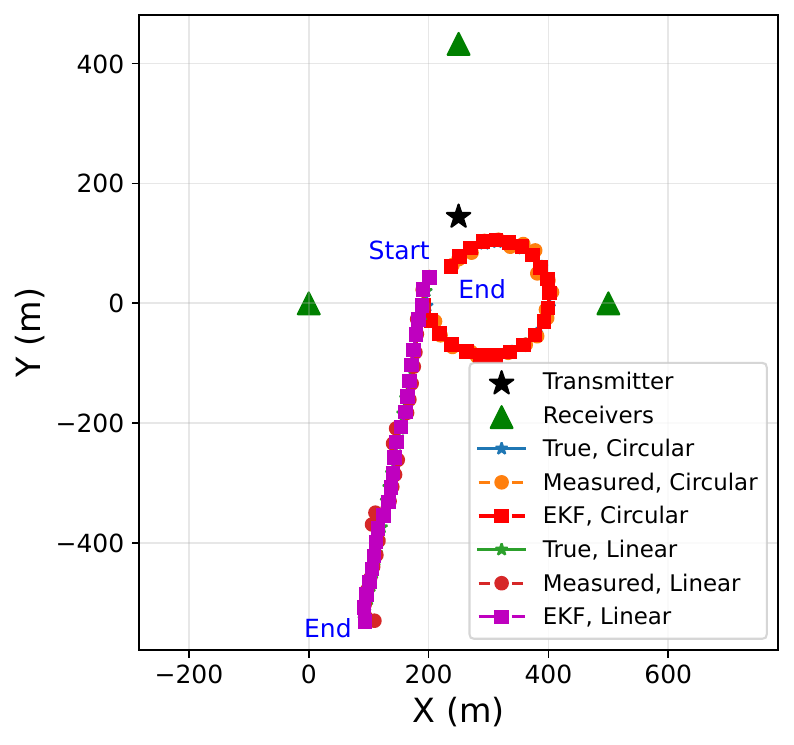}
  \caption{Tracks for different filters and motion models.}
  \label{fig:tracking_linear_circular_EKF}
  \vspace{-0.35cm}
\end{figure}

Table~\ref{tab:ekf_results} shows the total measurement and filtering errors under linear and circular motion profiles using the standard KF and EKF. The EKF can effectively track both linear and circular motions, while the standard KF yields slightly more accurate results for the linear trajectory but fails in closely tracking nonlinear (circular) motions. Fig.~\ref{fig:tracking_linear_circular_EKF} illustrates the tracks achieved by the EKF under both motion profiles in comparison with measurement results for positioning.

\begin{table}[ht!]
\centering
\caption{Total position errors in tracking.}
\label{tab:ekf_results}
\setlength{\tabcolsep}{3pt}
\begin{tabular}{llrr}
\toprule
Filter       & Motion    & Measurement error (m) & Filtering error (m) \\
\midrule
Standard KF    & Linear    & 144.90               &  97.25             \\
   & Circular    & 138.16             &   818.76            \\
Extended KF  & Linear    & 144.90               &  101.39             \\
             & Circular  & 138.16               &  63.89             \\
\bottomrule
\end{tabular}
\end{table}

\section{Conclusion} \label{sec:conclusion}
In this paper, we have presented MULTI-SCOUT that repurposes standard 5G PRS waveforms for high-accuracy multistatic sensing and tracking of moving targets. By extracting coherent range and Doppler measurements
across distributed receivers,
MULTI-SCOUT delivers fine-grained 2D and 3D 
positioning, clock calibration for time synchronization, and target association across distributed receivers.
The position estimates are then processed by standard KF or EKF for target tracking under different motion models. Overall, MULTI-SCOUT provides high-fidelity moving target detection, positioning, and tracking using the existing 5G PRS signals without relying on additional sensing capabilities.

\bibliographystyle{IEEEtran}
\bibliography{references}
\end{document}